  \documentstyle[11pt,aaspp4]{article}  % for astro-ph

\def\lap{\hbox{${_{\displaystyle<}\atop^{\displaystyle\sim}}$}}

\def\Mesz{M\'esz\'aros}

\begin{document} 
\title{Pulse Width Evolution in Gamma-Ray Bursts:  Evidence
for Internal Shocks}
 \author{E.~Ramirez-Ruiz$^{1,2}$ and E.~E.~Fenimore$^{1}$}  
\affil{$^1$MS D436, Los Alamos National Laboratory, Los Alamos, NM 87545}
\affil{$^2$Institute of Astronomy, Madingley Road, Cambridge CB3 0HA UK}

\begin{abstract}
Many cosmological models of GRBs envision the energy source to be a
cataclysmic stellar event leading to a relativistically expanding
fireball. Particles are thought
to be accelerated at shocks and produce nonthermal radiation. The
highly variable temporal structure observed in most GRBs has
significantly constrained models.  By using different methods of
statistical analysis in the time domain we find  that the width of the
large amplitude  pulses in GRB time histories remains remarkably
constant throughout the classic GRB phase. This is also true for
small amplitude pulses. However, small and large pulses do not have
the same pulse width within a single time history.  We find a
quantitative relationship between pulse amplitude and pulse width:  the
smaller amplitude peaks tend to be wider, with the pulse width following
a power law with 
an index of $\sim$ -2.8.
Internal shocks simulated by randomly selecting the Lorentz factor and
energy per shell are consistent with  a power law relationship. 
This
is strong quantitative evidence that GRBs are indeed caused by internal
shocks.
The dependency of the width-vs.-intensity relationship on the maximum
Lorentz factor provides a way to estimate that elusive parameter. Our
observed power law index indicates that $\Gamma_{\rm max}$ is $\lap 10^3$.
 We also interpret the narrowness of the pulse width distribution
as indicating that the emission which occurs when one shell overtakes
another is produced over a small range of distances from the central
site.         
\end{abstract}

\keywords{gamma-rays: bursts - internal shocks}

\newpage

\section{Introduction}
The cosmological origin of GRBs, established as a result of optical
follow-up observations of fading X-ray counterparts to GRBs
(\cite{cos97}), requires
an extraordinarily large amount of energy to flood the
entire universe with gamma rays ($10^{52}-10^{54}$ erg).
The source of this energy is assumed to be a
cataclysmic event (neutron
star-neutron star merger, neutron star-black hole merger, or the formation
of a black
hole). The lack of apparent
photon-photon attenuation of high energy photons implies substantial
bulk relativistic motion. The relativistic shell must have a high
Lorentz factor, $\Gamma = (1-\beta^{2})^{-1/2}$, 
on the order of $10^{2}$ to $10^{3}$.

A growing consensus is that a central site releases energy in the
form of a wind or multiple shells over a period of time commensurate
with the observed duration of GRBs (\cite{rm94}). Each subpeak in the
GRB is the result of a separate explosive event at the central
site. General kinematic considerations impose constraints on the
temporal structure produced when the energy of a relativistic shell is
converted to radiation.

The purpose of this paper is to
analyze the time histories of many GRBs to uncover the temporal
evolution of the pulse width. In an earlier report (\cite{rf99}) we
found no  significant change in the average peak width in long
bursts. Here, we analyze both long and short bursts in greater
detail, as well as small and large amplitude pulses in individual
bursts, and compare the results to internal shock models.

\section{Observations}

\subsection{Temporal Evolution of the Average Pulse Width}

Gamma-ray burst temporal profiles are enormously varied. Many
bursts have a highly variable temporal profile  with a time scale
variability that is significantly shorter than the overall duration.
Our aim is to  characterize  and measure the pulse shape as a function
of arrival time. We will use the aligned peak method which measures the
average pulse temporal structure by aligning the largest peak of each
burst (\cite{mi93}).

The Burst and Transient Source Experiment (BATSE) catalog provides
durations called $T_{90}$ (\cite{meg96}), where $T_{90}$ is the time
which contains 90\% of the counts.  For the purpose of our analysis, we
used two sets of bursts from
the BATSE 4B Catalog that were brighter than 5 photons s$^{-1}$
cm$^{-2}$ (256 ms peak flux measured in the 50-300 keV energy range)
and with a 64ms temporal resolution. 
The first set used all 53 bursts that were longer than 20s,
and the second set used all 23 bursts that were shorter than 20s.  
Each burst must have at least one peak, as
determined by a peak-finding algorithm (similar to \cite{hli96}), in
each  third of its duration. The largest peak in each third was
normalized to unity and shifted in time, bringing the largest peaks of 
all bursts into common alignment. This method was applied in each third
of the duration of the bursts. Thus, we  obtained one
averaged pulse shape, $I(t)$, for each third of
the bursts (as shown in Figure \ref{peakalign}a for the long duration
bursts and in Figure \ref{peakalign}b for the short duration
bursts).The average width is notably identical in each 1/3 of $T_{90}$ in both
long and short bursts. We estimate the differential spread, $S$, to be
$\lesssim$ 1\% for the long duration bursts and $\lesssim$ 5\% for the
short duration bursts.  

The values $I(t)$ along the aligned timescale
represent the average level of the emissivity of all contributing
sources aligned at their primary peaks and thus the general character
of the emission evolution of GRBs (see \cite{mi93} for details). To
resolve the true differences between the timescales in GRB pulses, one
has to find the appropriate temporal  correspondences in order to align
the events, despite their probably different time histories. However,
such a correspondence seems to exist because each burst has a specific
moment, namely, the highest peak of the time history, which may be
regarded as a physically unique reference moment. Furthermore, the
highest peak is also where the highest signal-to-noise ratio is
observed. The selection of a high brightness sample (5 photons s$^{-1}$
cm$^{-2}$ in this case) is appropriate  in order to avoid
systematic effects that might change the observed time histories with different
statistics. The time histories of dim events would be more randomized
by fluctuations than the time histories of bright bursts. Using other
GRB samples with a high signal-to-noise ratio ($\gtrsim$ 3 photons
s$^{-1}$ cm$^{-2}$) gives similar results.

Figure \ref{peakalign} shows that the pulse width does not increase with
time. It could be argued that the peak alignment method is
uncertain because it only reflects the temporal evolution of the
largest pulse width in the time histories. Thus, in the following
section, we expand our analysis to individual pulses in GRB time
histories.

\subsection{Average Temporal Evolution of the Pulse Width}
  
The substantial overlap of the temporal structures in
the burst have made the study of individual pulses somewhat
difficult. An excellent
analysis has been provided by \cite{jn96}, who examined the
temporal structure of bright GRBs by fitting
time histories with pulses. The time histories were fit until all
structure was accounted for within the statistics, thus, they
effectively deconvolved the time history into all of the constituent
pulses. From the set of pulses that they analyzed, we
used the 28 bursts that have five or more
fitted pulses (in the 55 keV - 115 keV BATSE channel) within their
$T_{90}$ duration. There was a total of 387 pulses in those 28
bursts. We obtain the Full Width Half Maximum (FWHM) from the pulse
shape parameters found by \cite{jn96}. To find the average pulse width 
as a function of time, we first normalized the FWHM of
each peak, within a burst, to the average FWHM of that burst.  
The purpose of such normalization is so that no one burst is allowed
to dominate the pulse width average. Second, we normalized all
the pulse amplitudes to the average amplitude. This is required in
order to differentiate intrinsically large and small pulses in all bursts
despite
the total net counts. Figure \ref{widvstime} shows the average pulse
width, ${W \over <W>}$, as a function of  temporal position in
the time history. 
The filled
symbols give the average normalized width of the pulses that
have a normalized amplitude, ${A \over <A>}$,  greater than 1.0, while
the open symbols show ${W \over <W>}$ for the pulses with a normalized
amplitude less than 1.0.  
Each group has about
180 pulses. The resulting average (in both samples) appears to be 
fairly constant in time. One cannot determine strict
error bars because the uncertainties are not due to counting statistics
(which, after all, are very good, since we are adding together $\sim$
180 pulses ). Rather, the fluctuations are
due to the way  in which the various peaks add together. We used a
linear fit to search for a trend. The resulting average temporal evolution of
the pulse width  is remarkably constant for both samples: 
\begin{equation}
{W \over <W>}=  0.82 -0.01 {T \over T_{90} }~~~~~~~~~~~{\rm
if}~~{A \over <A>}~>~1.0
$$
$$
{}~~~~~~~~   =  1.28 -0.02 {T \over T_{90} }~~~~~~~~~~~{\rm if}~~{A \over
<A>}~<~1.0~.
\label{WIDTIME}
\end{equation}
These curves are shown in Figure \ref{widvstime}  as dotted lines.

A visual inspection of the pulses fitted to gamma ray bursts by
\cite{jn96} shows that the low amplitude pulses (in a single burst)
tend to be wider although their shape may not be well
determined. This is due to the fact that the actual temporal profile
may contain ``hidden'' pulses which are not easy to deconvolve but
contribute to the total emission. Furthermore, there are
pulses that may overlap their neighbors and the pulse model is not
sufficiently detailed to represent all of the individual emission
events. Thus, larger pulses are much more succesfully deconvolved than
the smaller amplitude ones. It is more difficult to  conclude that the average
temporal
evolution of the pulse width in small amplitude pulses is as
constant over $T_{90}$ as that of the large amplitude pulses for two reasons.
First, the standard
deviation of the distributions of pulse
width values for small peaks  is
$\sim$ 1.7 - 2.2 times greater than that found for
the analysis of large amplitude pulses. Second, the {\rm linear
correlation coefficient} of the linear fit to the large amplitude
pulses is $\sim$ 1.12 times greater than the one found in the linear
fit to the small amplitude pulses. Nevertheless, the small pulses show the
same consistency in pulse widths as the large pulses.

This analysis of individual time histories agrees with what was found for
the
evolution of the average pulse structure for large peaks. Individual
bursts show that larger peaks have about the same width
at the beginning of the burst as near the end of the burst with a
rather small variation. This is also true for smaller pulses. However,
as we show in the next section,
small and large pulses do not have the same pulse width within a
single profile.

\subsection{Pulse Width as a Function of Amplitude}
GRBs are very diverse, with time histories ranging from as short as
50 ms to longer than $10^{3}$ s. The long bursts often have very
complex temporal structure with many subpeaks. The process that
produces the peaks has a random nature, and the peaks that are
produced vary substantially in amplitude. These pulses tend to be
wider as their amplitudes decrease, within a single profile. To
investigate the amplitude dependency of the pulse width, we used
the 28 bursts described in section 2.2. Each pulse
(in each profile) was normalized to the average amplitude found in that
burst. We selected four regions of normalized amplitudes: 0.1 - 0.3,
0.3 - 0.9, 0.9 - 1.5, 1.5 - 2.0. Each group has about 95 pulses. Figure
\ref{widvsint} 
shows the aligned average pulse shape for the four ranges of
normalized amplitudes. The pulse shape was calculated
based on the general pulse shape proposed by \cite{jn96}:

\begin{equation}
\label{NORRISSHAPE}
{\rm I(t)}= {\rm A~~exp} [-({|t-t_{peak}| \over \sigma_r})^{\nu}]~~~~
{\rm if}~~{\rm t~<~t_{peak}},
$$
$$
{}~~~~~~= {\rm A~~exp} [-({|t-t_{peak}| \over \sigma_d})^{\nu}]~~~~
{\rm if}~~{\rm t~>~t_{peak}},
\end{equation}
where $t_{peak}$ is the time of the pulse's maximum intensity ($A$);
$\sigma_r$ and $\sigma_d$ are the rise and decay time constants,
respectively; and $\nu$ is a measure of pulse sharpness, which
\cite{jn96} refer to as ``peakedness''.

Each pulse,
in each range of amplitude, was normalized to unity and then shifted
in time, bringing the center of all pulses into common alignment. Note
that the smaller peaks are wider than the larger peaks (see Fig.
\ref{widvsint}).
We  characterized the amplitude dependency of the pulse width in
GRB time histories using the FWHM of the aligned average pulse
shapes in Figure \ref{widvsint}. The open diamonds in Figure
\ref{widvsintfit}  are the
widths, $W$,
of each aligned average pulse shape measured at the half maximum. We
have fitted a power law and an exponential function to the points. The best-fit
power law is ${A \over <A>} \sim [W_{FWHM}]^{-2.8}$ (The exponential fits
had $\chi^{2}$ values that were 1.4 times larger). The power-law function is
shown in Figure \ref{widvsintfit} as a dotted line. This a robust result.
Using the
width at other values of the average pulse shape  gives
similar results. One thing that is not clear in our formulation is at
what normalized amplitude to place the points. We have placed them at
the mid-point of the  selected amplitude ranges. 
If we were to use the average point of all the
normalized amplitudes in each selected range, the result is still a power law:
${A \over <A>} \sim [W_{FWHM}]^{-3.0}$. 

In summary, we find that the aligned average pulse shape can measure
the amplitude dependency of the pulse width. The dependency is a power
law in pulse width with an index that is between -2.8 and
-3.0, depending on how it is measured.
       
Some limitations are necessarily inherent in our approach and
selection of data. The conclusions we reach are based on measurements
of a subset of the bursts detected by BATSE. We analyze the pulses
deconvolved by \cite{jn96} from a subset of relatively bright bursts
with 64 ms temporal resolution. Analysis of pulses in shorter bursts
using different data with much higher resolution will be the subject
of another paper. In a multipeaked
event, all peaks would be seen if the burst is intense, whereas some
peaks might be missed in a weak version of the event owing to the
decrease of the signal-to-noise ratio. Moreover, smaller peaks of
dimmer events might be missed owing to the absence of triggering of
the instrument at those peaks. These effects might lead to a
systematic decrease of the average number of peaks and/or to a
decrease of estimated burst duration with decreasing burst
intensity. Although our approach utilizes a
sufficient number of pulses to represent adequately the temporal
profile of a certain burst, our inferences concerning pulse shape are
drawn from those fitted pulses which do not overlap, as estimated by
the relative amplitudes of two pulses and the intervening
minimum. Several mutually reinforcing trends have been found by
\cite{jn96} in the analysis of the same sample. Thus, supporting the
validity of our results.    

From a phenomenological point of view, it
has not been clear what the fundamental ``event''  in gamma-ray
bursts is. The premise of our work has been that pulses are the basic unit in
bursts. The relationship between pulse width and intensity supports
this hypothesis. However, there may be other components in
bursts, undefined by our approach, including long smooth structures at
lower energies or  very short spiky features at higher energies, which
might represent distinct physical processes from the ones that are
responsible for pulse emission. Evidence for a separate
emission component, similar to those of the afterglows at lower
energies, has been clearly found in some GRB light curves
(\cite{gib99}). These observations may indicate that some sources
display a continued activity (at a variable level). 

\section{Pulses From Internal Shocks}

Internal shocks occur when the relativistic ejecta from the central
site are not moving uniformly. If some inner shell
moves faster than an outer one ($\Gamma_i > \Gamma_j$) it will
overtake the slower at a radius $R_{c}$.  The two
shells will merge to form a single one with a Lorentz factor $\Gamma_{ij}$. The
emitted radiation from each collision will be observed as a single
pulse in the time history (\cite{ps97}, \cite{sp95}).
Several groups have modeled this process by randomly selecting the initial
conditions at the central site (\cite{kob97}, \cite{dm98}, 
\cite{fr99}).
We will compare two aspects of these internal shock models to the pulse
evolution studied in this paper: the trend for smaller pulses to be wider
and the narrowness of the pulse width distribution.

\subsection{Pulse Width vs. Intensity}

We have simulated internal shocks as described in \cite{fr99}. In the
notation of that paper, we have set the maximum initial energy per
shell, $E_{\rm max}$, 
to be $10^{53.5}$ erg, the maximum thickness to be 0.2 lt-s, and the
ambient density to be 1.0 cm$^{-3}$. We generated about 1.4 shells per s.
Nine values of the maximum Lorentz factor, $\Gamma_{\rm max}$, were
simulated from $10^{2.5}$ to $10^{4.5}$. The minimum Lorentz factor,
$\Gamma_{\rm min}$, was 100. We took the resulting pulses and determined
the peak intensity (assuming 0.064 s samples) and the FWHM. Figure
\ref{iswidvsint}a shows the distribution of pulse widths and intensities if
$\Gamma_{\rm max}$ if $10^{2.5}$ and \ref{iswidvsint}b is for $\Gamma_{\rm
max}$ is $10^{4.5}$. The solid line is a power law with the index 
determined from the
observations (i.e., from Fig. \ref{widvsintfit}). Internal shocks show a
trend that smaller pulses are wider. Indeed, one can estimate $\Gamma_{\rm
max}$ by measuring the index of the width vs. intensity distribution.
By running models with a variety of values of $\Gamma_{\rm max}$, we have
found that the index is $\sim -5.25+0.975 {\rm log}_{10}(\Gamma_{\rm
max})$.
Our observed index of $\sim -2.8$ (from  Fig. \ref{widvsintfit}) indicates
that $\Gamma_{\rm max}$ is  $\lap 10^3$.

\subsection{Pulse Width as an indicator of $R_c$}

A shell that coasts without emitting
photons and then emits for a short period of time produces a pulse with a
rise time related to the time the shell emits and a decay dominated by
curvature effects (\cite{fmn96n}).  In the internal shock model,
the  shell emits for $\Delta t_{\rm cross}$, where  $\Delta t_{\rm
cross}$ is the time it takes the reverse shock to  cross the shell that is
catching up. Following \cite{kob97}, $\Delta t_{\rm cross} =
l_j/(\beta_j-\beta_{rs})$, where $l_j$ is the width of the rapid shell
($\beta_j$).

To calculate the observed pulse shape, one needs to
combine Doppler beaming with the volume of material that can
contribute at time $T$. Following \cite{fr99} and \cite{chiphunt}, the
resulting pulse shape is 
\begin{eqnarray}
V(T) &=&0\hspace{325pt}{\rm if}\ T<0  \nonumber \\
&=&\psi \frac{(R_c+2\Gamma_{ij}^2cT)^{\alpha +3}
-R_c^{\alpha +3}}{(R_c+2\Gamma_{ij}^2cT)^{\alpha +1}}
\hspace{122pt}{\rm if}\ 0<2\Gamma_{ij}^2T<\Delta t_{\rm cross}
\label{CHIPENVEL} \\
&=&\psi \frac{(R_c+\Delta t_{\rm cross})^{\alpha +3}-
R_c^{\alpha +3}}{(R_c+2\Gamma_{ij}^2cT)^{\alpha+1}}
\hspace{147pt}
{\rm if}\ 2\Gamma_{ij}^2T>\Delta t_{\rm cross}
\nonumber
\end{eqnarray}
where $\psi$ is a constant, $T$ is measured from the start of the
pulse and $\alpha$ ($\sim$ 1.5) is the power-law index of the
rest-frame photon number spectrum. The amplitude, $\psi$, depends
on the amount of energy converted to gamma rays in a given collision.

Figure \ref{rcvswid} shows the FWHM obtained from equation \ref{CHIPENVEL} 
(assuming
that
$l_j$ = 1 light second and ${\Gamma_i \over \Gamma_j}$=10) as a function of the
radius of emission, $R_c$, and the Lorentz factor of the resulting shell,  
$\Gamma_{ij}$. Note that a wide range of widths map into a narrow
range of radii. In the internal shock scenario, the observed temporal
structure reflects directly the activity of the inner engine. In
Figure \ref{dynamicrc} we show the distribution of radii of emission
using the FWHM calculated by equation \ref{NORRISSHAPE} 
for the parameters provided by
\cite{jn96n}. The FWHM is used with Figure \ref{rcvswid} to find a radius
for each of the 387 pulses.
The radius of emission is normalized by $\Gamma_{ij}^{2}$, and, since the
curves in Figure \ref{rcvswid} are self-similar, all values of
$\Gamma_{ij}$ give the same distribution when divided by
$\Gamma_{ij}^{2}$.
 We define the radius spread,
${\Delta R_c \over R_c}$, to be the ratio between the center and the
standard deviation of the distribution of the radius of emission.
This distribution shows that, if the spread of values of the Lorentz
factors of the shells ($\Gamma_{ij}$) is small, the dynamical
range of the radii of emission is also small: $\Delta R_c \sim 0.07 R_c $. The
multiple-peaked time histories  in the BATSE catalog reveal that the
dynamical ranges in observed timescales within cosmic gamma-ray bursts
(GRBs) are very large (see \cite{jn96}). For example, the total event
durations range from 10 ms to 1000 s, with a dynamical range of almost
$10^{5}$. Thus, the small variation in the values of the pulse width 
and radius spread 
parameter is remarkable.

In the internal shock scenario, the observed temporal structure
reflects directly the activity of the inner engine. This engine must
operate for a long duration, up to hundreds of seconds in some cases, and
it must produce a highly variable wind to form shells that
radiate. If the spread of values of the Lorentz factors 
($\Gamma_{ij}$) is small, the range of radius is $\Delta
R_c \sim 0.07 R_c$. The arrival time of the pulses at a detector such
as BATSE has a one-to-one
relationship with the time the shell was created at the central site. The
time of arrival, $T_{\rm toa}$, is $t_{ij}-R_c/c$ where $t_{ij}$ is the
time of the collision. But $t_{ij}-R_c/c$ is roughly $t_{oi}$, the time
the shell was produced at the central site and
is not  dependent on other parameters such as $R_c$ or the time of
the
collision (see, for example, Eq. 5 in \cite{fr99}). Thus, internal shocks
also explain why the pulse width tends to be constant throughout the
burst: the time of arrival in the time history is effectively just the
time of generation of the pulses at the central site and is not related to
the conditions or parameters of the collision.

\section{Summary}

We calculated the temporal evolution of the pulse width in gamma ray
bursts. We found that the average aligned pulse width is a
universal function that can measure the timescale of the largest
pulses in the burst. For long and short bursts we found that the
average aligned pulse width undergoes no significant change during the
gamma-ray phase (see Fig. \ref{peakalign}). The analysis of individual time
histories
agrees with what was found in the average aligned method. Individual
bursts typically have no time evolution of the width of the largest
pulses. This is also true for small pulses (see Fig. \ref{widvstime}).
However, in a
time history, the smallest amplitude peaks tend to be wider (see
Fig. \ref{widvsint}). The dependency, as shown in Figure
\ref{widvsintfit}, is a power law
in a
amplitude with an index that is between -2.8 and -3.0, depending on
how it is measured.

We have found that internal shocks can
explain most of these characteristics.
The time of arrival of a pulse is not related to the collision parameters
so internal shocks can produce pulses that have the same characteristics
at the beginning as at the end. Internal shocks produce pulses that
are wider for smaller intensities. If the maximum
$\Gamma$ is $\lap 10^3$ the observed distribution (Fig.
\ref{widvsintfit}) is similar to the simulated distribution (Fig.
\ref{iswidvsint}). For such low values of $\Gamma$, deceleration is
usually not important and the simulated time histories do not have pulses
that get progressively wider (see \cite{fr99}). This is consistent with
the analysis of this paper which did not find progressively wider pulses,
although such pulses might have been missed because it is difficult
to deconvolve many overlapping small pulses. Without substantial
deceleration, the efficiency for converting bulk motion into radiation is
$\lap$ 25\% (\cite{fr99}). 

In the internal shock scenario, the
temporal structure  directly reflects the temporal behavior of the
inner engine that drives the GRB. The  pulse width
gives  information about the radius of colliding shells.
Figure \ref{rcvswid} shows that a wide range of widths maps into a narrow
range of radii (see Fig. \ref{dynamicrc}).

\begin{acknowledgements}
We thank Jay Norris for providing the pulse-fit parameters. 
\end{acknowledgements}

\clearpage

\figcaption[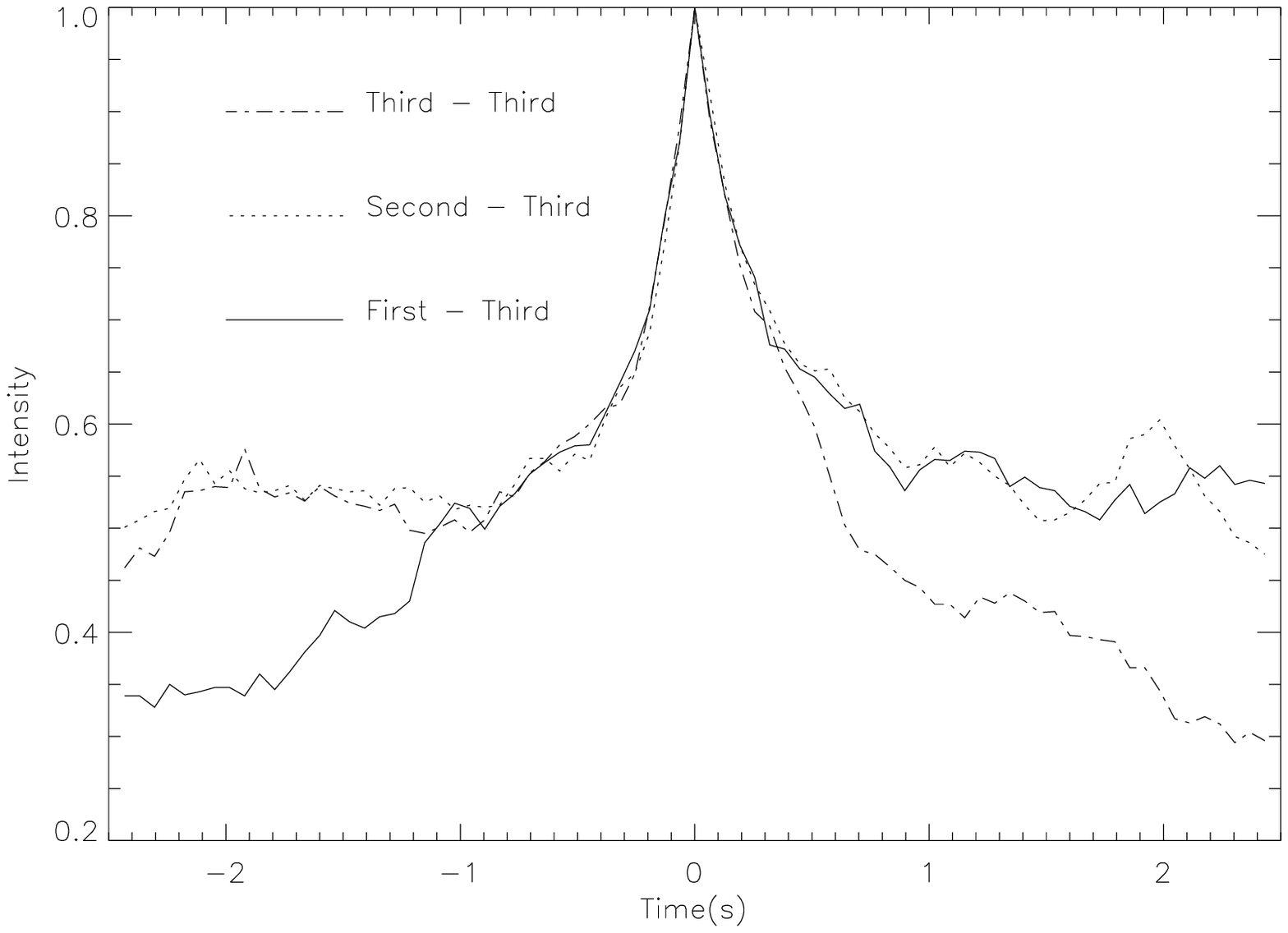]{Average peak alignment of gamma-ray
emission from bright
BATSE bursts. The largest peak in each third of the burst duration (i.e.
$T_{90}$) is aligned and averaged.  The three curves show
the average pulse shape for the largest peak in the first third, second
third, and last third of the bursts. 
\hfill\break
(a)Average pulse shape from 53 bright BATSE bursts with durations
longer than 20 seconds. The difference in the average of the brightest
pulse in each third is about 1\%.
\hfill\break
(b)Average pulse shape from 23 bright BATSE bursts with durations
shorter than 20 seconds. The difference in the average of the brightest
pulse in each third is less than 5\%.
We find a lack of temporal
evolution of the width of the bright pulses over most of $T_{90}$.
In addition, the average width is nearly identical in short
and long bursts.
% {\rm NAMEFIG:peakalign}
\label{peakalign}
}

\figcaption[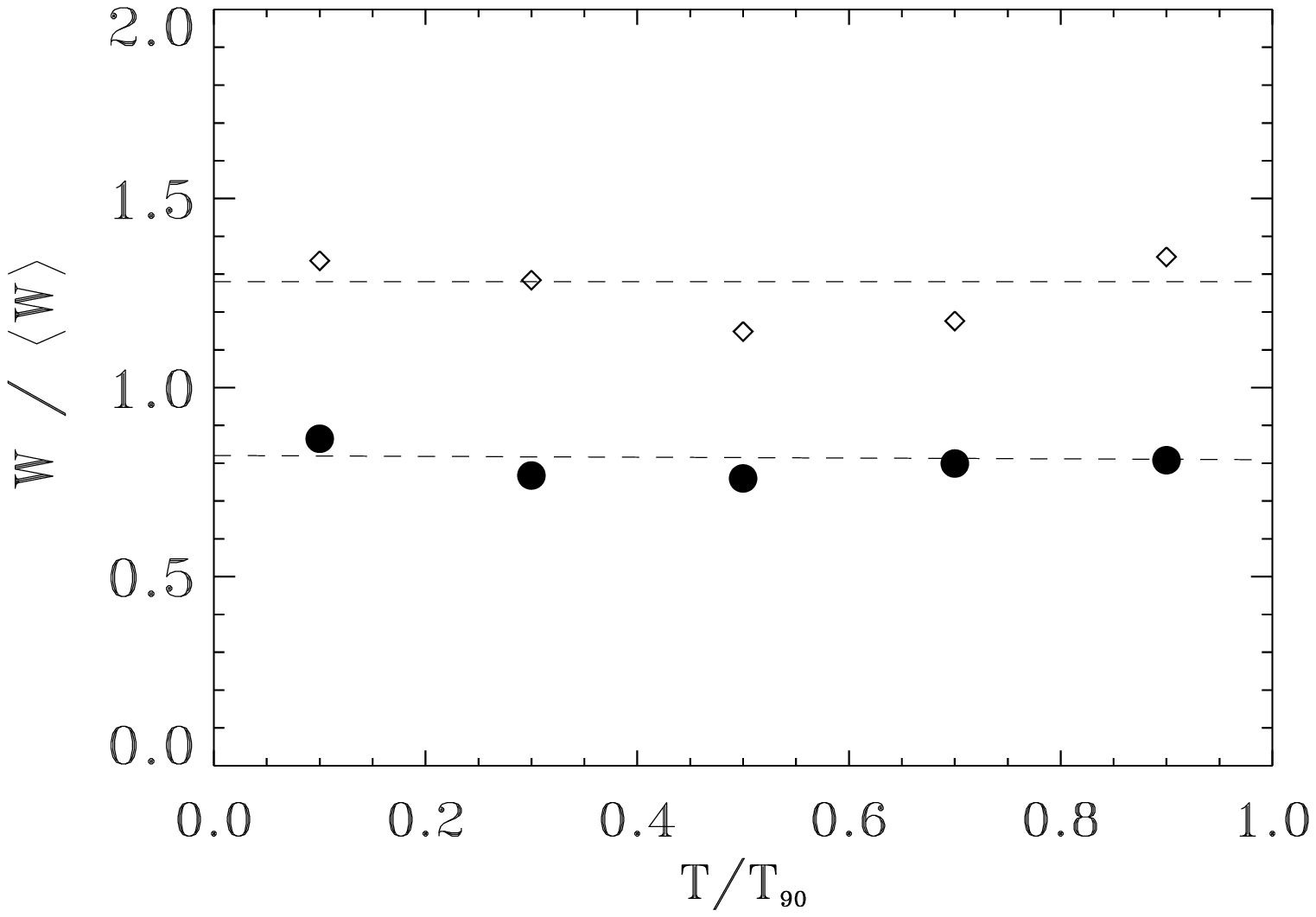]{
The evolution of pulse width with time.  We  sort the pulses into
five time ranges spanning $T_{90}$. We further sort the pulses into bright
peaks (amplitudes greater than the average for each burst, solid circles)
and dim peaks
(amplitudes less than the average, open triangles). Whereas Figure
\ref{peakalign} is based on the brightest peaks, all 387 pulses from 28
bursts are included in this figure,  divided roughly equally between
bright and dim.
For each group, we find the average of the  observed width normalized to
the average width for
that burst ($A/<A>$). The pattern that emerges is that both large and
small peaks have  remarkably constant widths throughout the duration
of 
the bursts but the dim peaks tend to be wider than the bright peaks. 
The dotted lines are linear fits to the data and the slopes are less than
2\%.
% {\rm NAMEFIG:widvstime}
\label{widvstime}
}

\figcaption[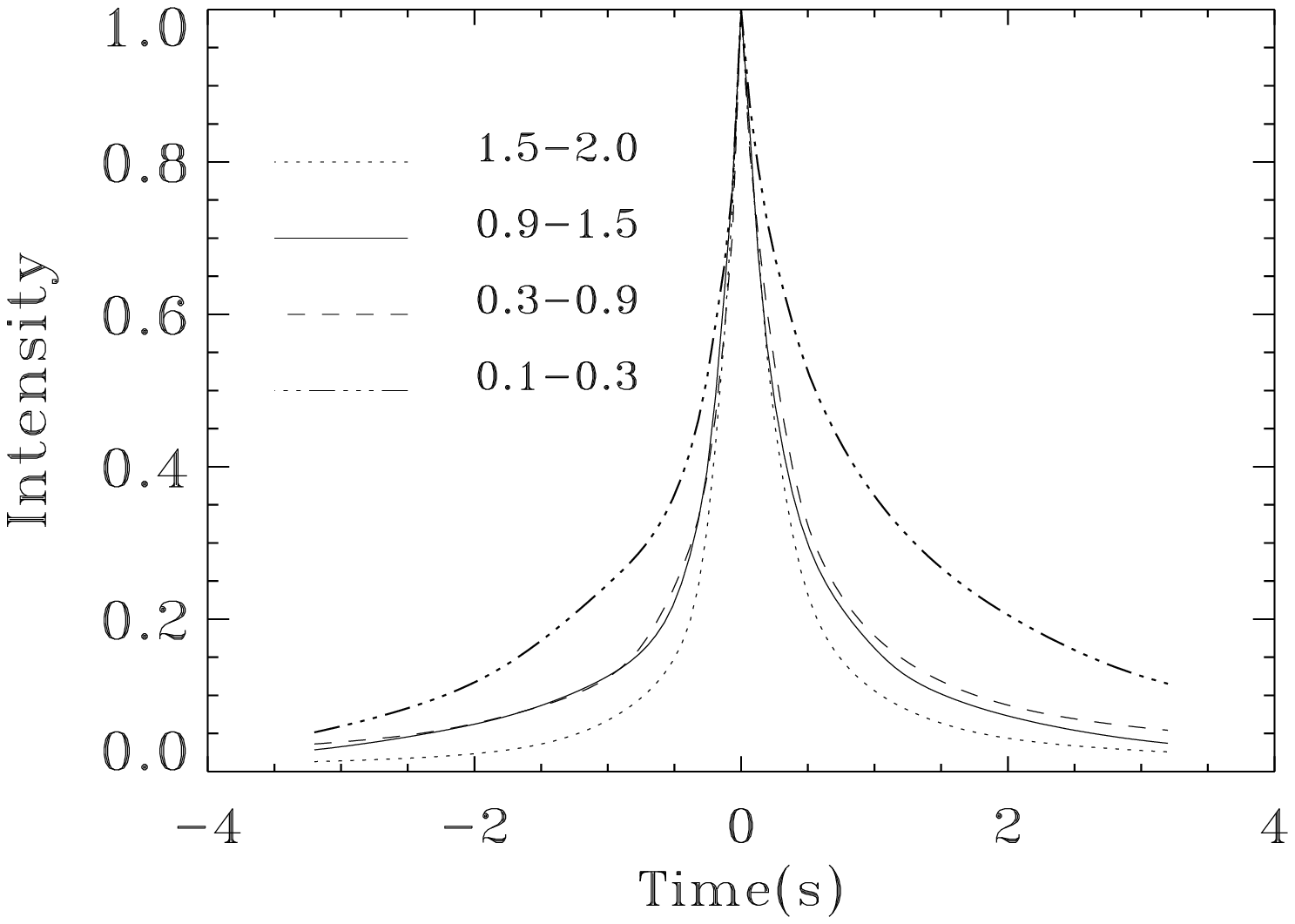]{
The average pulse profile as a function of relative intensity. We
normalize the intensity of each peak to the average intensity of each
burst. We then sort the 387 pulses from 28 bright bursts into four
relative intensity ranges spanning from 0.1 to 2.0. For
each group, we add the pulse shape Norris et al. (1996) found from
deconvolving the burst time
history into individual pulses. Thus, this could be considered an
aligned-pulse test where every pulse is aligned rather than just the
brightest peak. Ee do this for different relative intensities.
There is a clear trend: the larger peaks tend to be narrower.
% {\rm NAMEFIG:widvsint}
\label{widvsint}
}

\figcaption[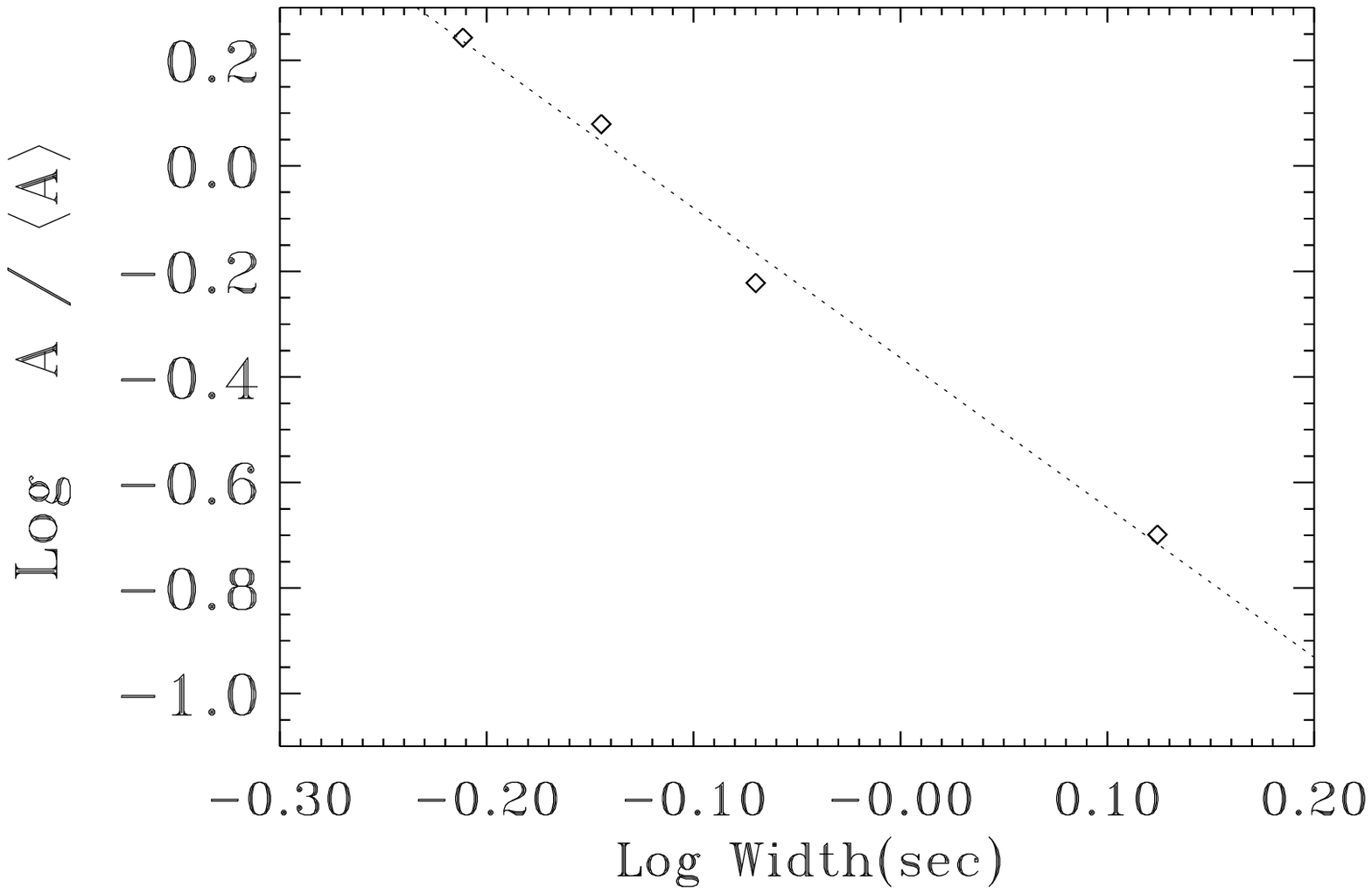]{
The intensity-width relationship. Short pulses are wider than large
pulses. From Figure \ref{widvsint}, the FWHM is a power law in intensity
with an index about -2.8 (dotted line).
% {\rm NAMEFIG:widvsintfit}
\label{widvsintfit}
}

\figcaption[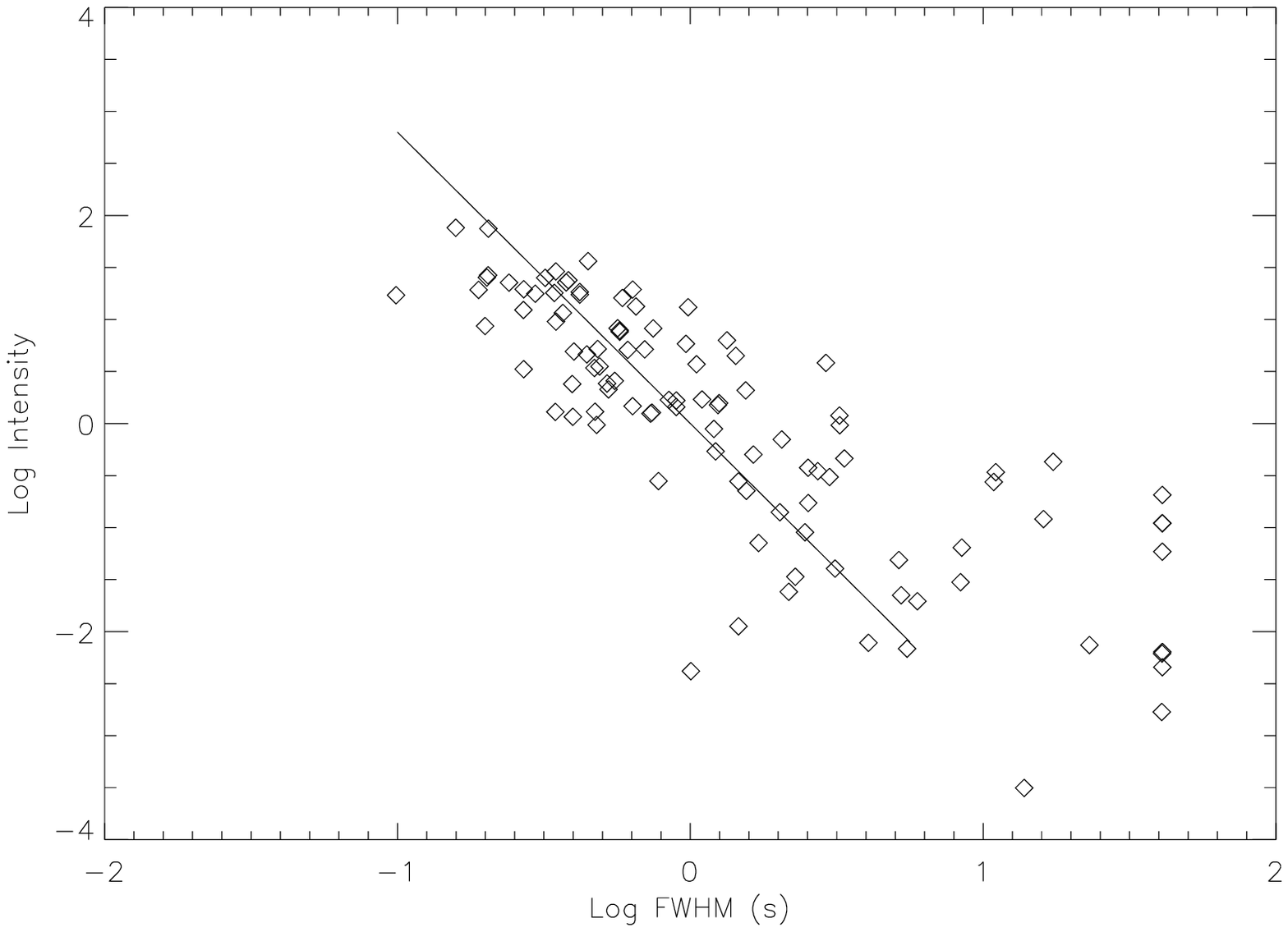]{
Pulse width vs. intensity from an internal shock model. Internal shocks
are modeled as in \cite{fr99} and the pulse width and intensity found for
each pulse.
\hfill\break
(a)Distribution if the maximum Lorentz factor is $10^{2.5}$. The solid
line
is the fit from Figure \ref{widvsintfit}; that is, a power law slope of
$\sim -2.8$.
\hfill\break
(a)Distribution if the maximum Lorentz factor is $10^{4.5}$. The
distribution is quite different from the observations (solid line) implying
that $\Gamma_{\rm max}$ is not as large as $10^{4.5}$.  
% {\rm NAMEFIG:iswidvsint}
\label{iswidvsint}
}

\figcaption[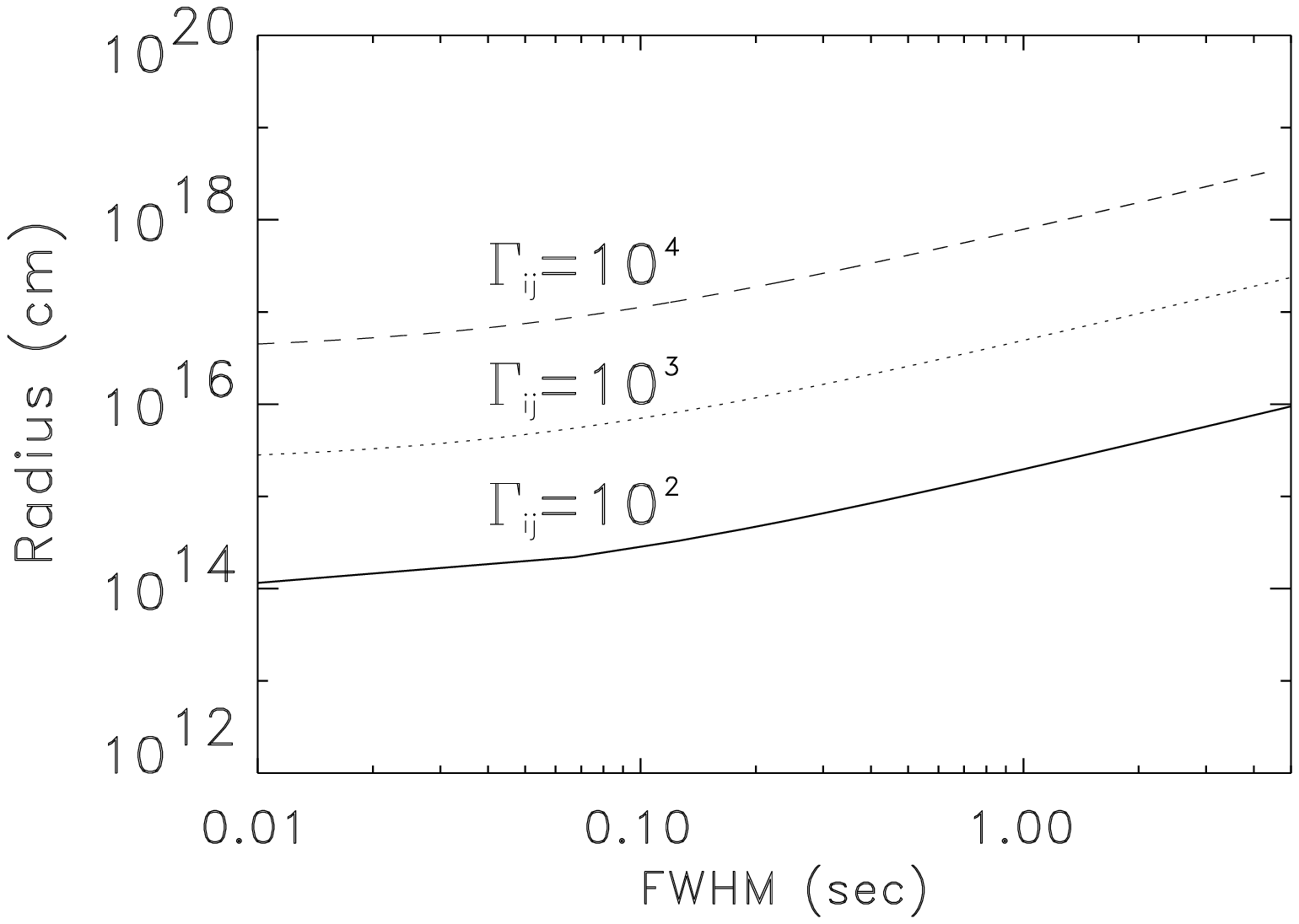]{
The expected pulse width as a function of the radius of colliding
shells. Equation \ref{CHIPENVEL} is used to estimate the pulse width from
the collision radius for various Lorentz factors ($l_j$= 1 light
second and ${\Gamma_i \over \Gamma_j}$=10).
Note that a wide range of widths maps into a narrow range of radii.
% {\rm NAMEFIG:rcvswid}
\label{rcvswid}
}

\figcaption[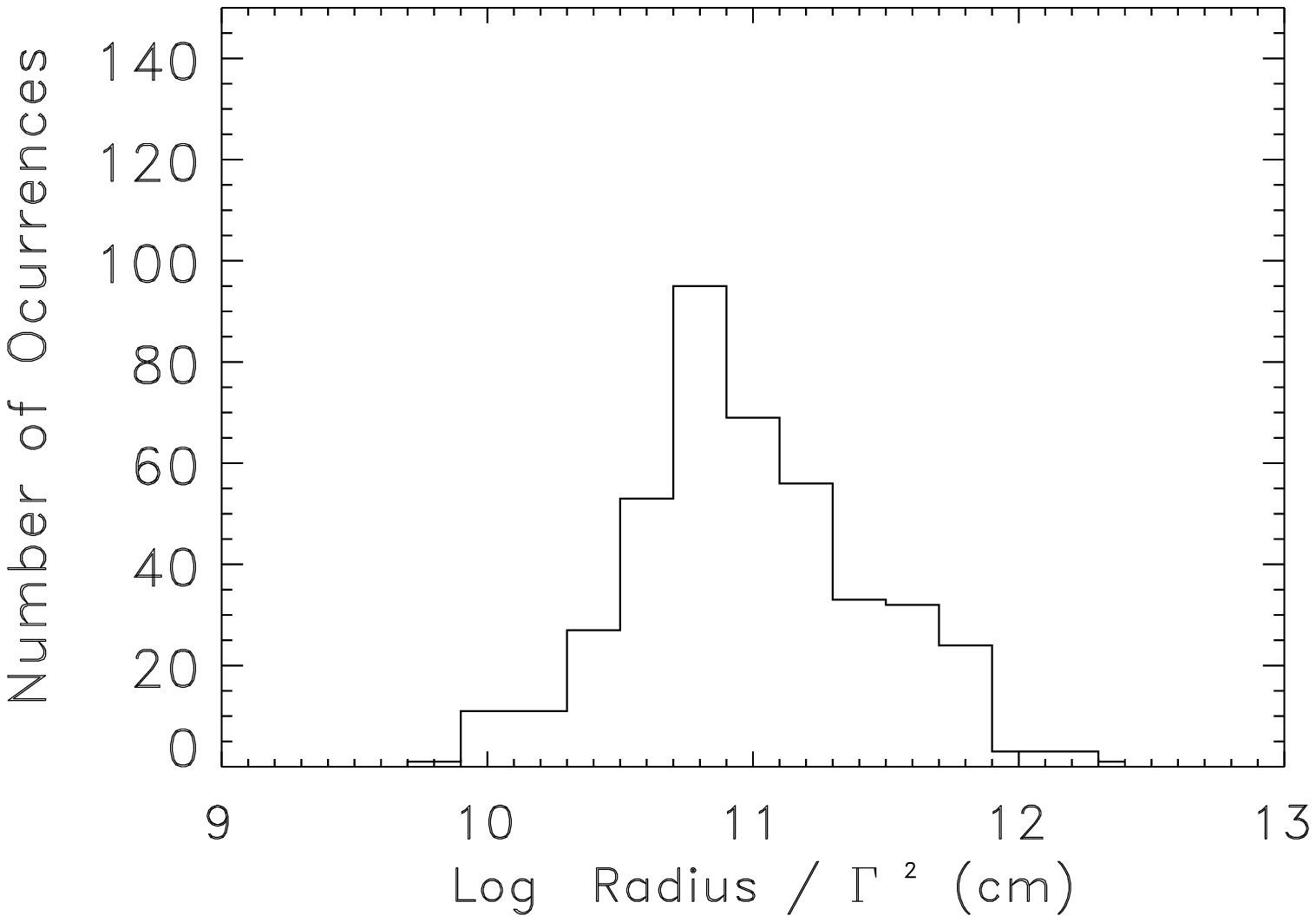]{
Distribution of collision radii based on observed pulse widths.
Equation \ref{CHIPENVEL} is used to calculate $R_c/\Gamma_{ij}^2$ for 387 pulses from
28 bright bursts. The width of the distribution is remarkably narrow,
$\sim$ 0.07\%.
% {\rm NAMEFIG:dynamicrc}
\label{dynamicrc}
}

%\end{document}

% Option 2.  The figure captions are printed on a caption page(s) as in 
% option 1.  The figures available as EPS files are then printed at the
% end of the document, one figure per page, using the \plotone command.
% If you wish to process this option then simply comment out the \end{document}
% just above these five lines. 

\clearpage

\centerline{Figure 1a}
\plotone{apj_pulses_fig1a.eps}
\clearpage
\centerline{Figure 1b}
\plotone{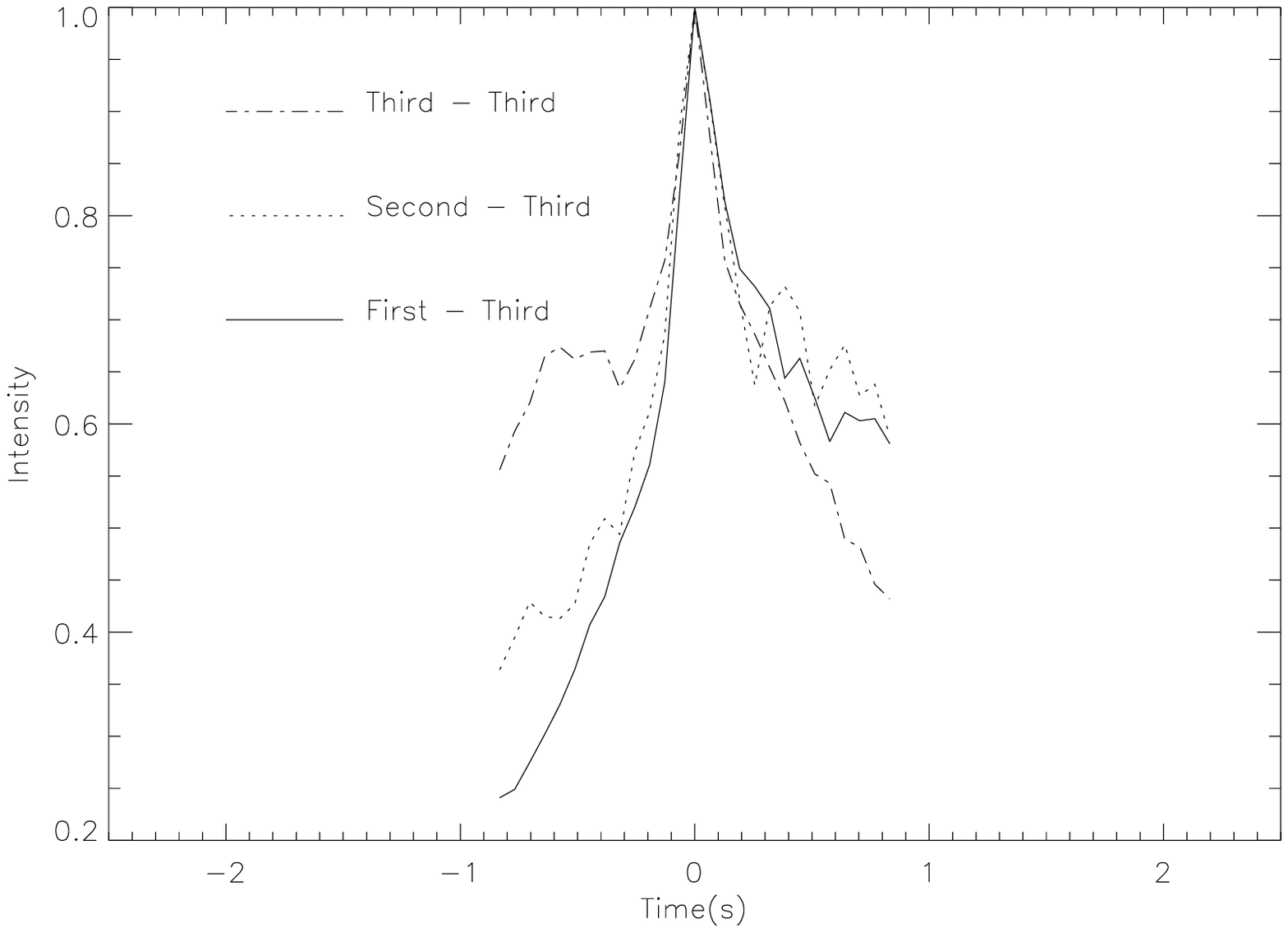}
\clearpage
\centerline{Figure 2}
\plotone{apj_pulses_fig2.eps}
\clearpage
\centerline{Figure 3}
\plotone{apj_pulses_fig3.eps}
\clearpage
\centerline{Figure 4}
\plotone{apj_pulses_fig4.eps}
\clearpage
\centerline{Figure 5a}
\plotone{apj_pulses_fig5a.eps}
\clearpage
\centerline{Figure 5b}
\plotone{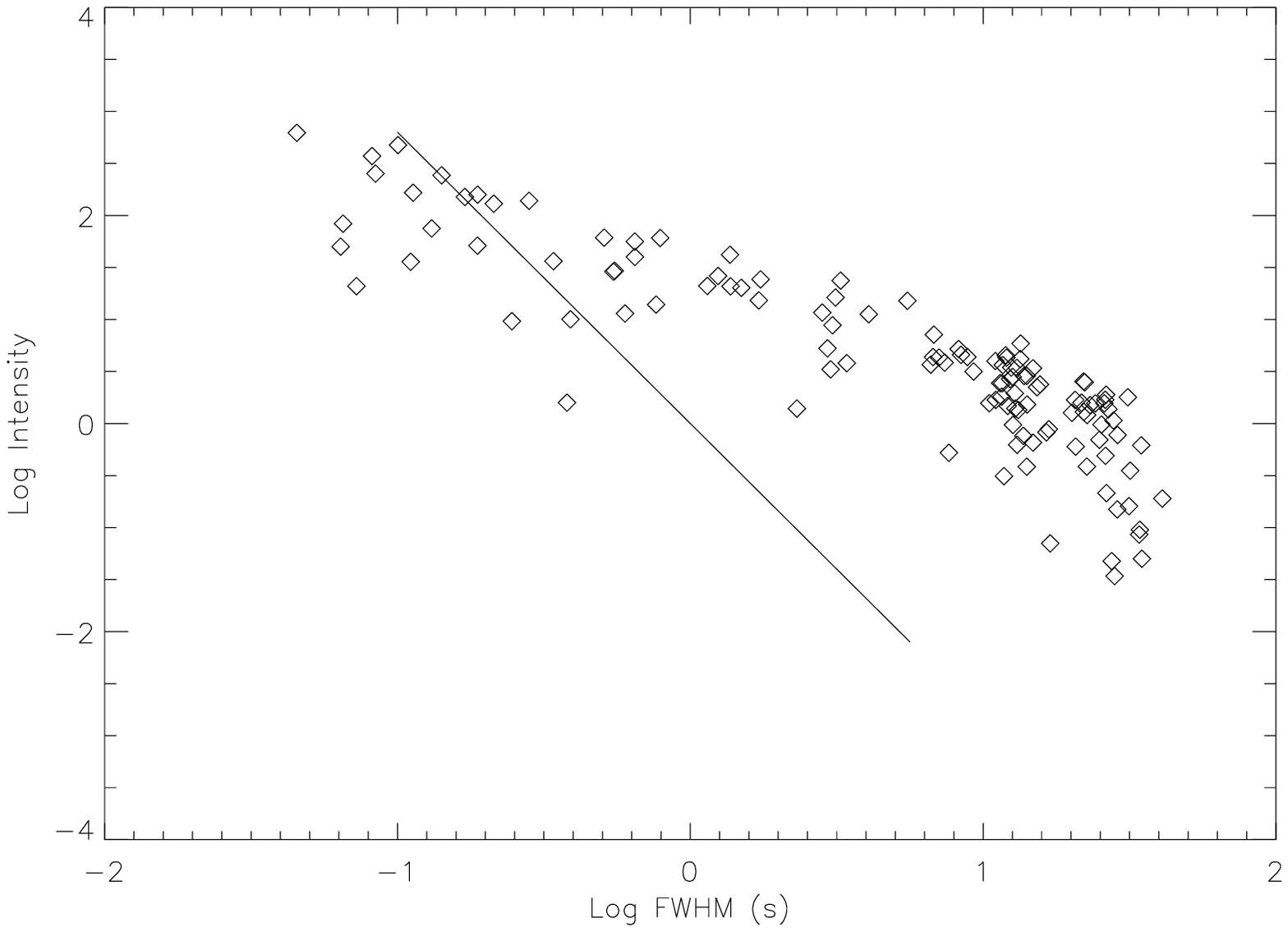}
\clearpage
\centerline{Figure 6}
\plotone{apj_pulses_fig6.eps}
\clearpage
\centerline{Figure 7}
\plotone{apj_pulses_fig7.eps}
\clearpage

\end{document}